\documentclass[12pt,showpacs]{article}
\usepackage{graphicx}
\begin{document}
\newcommand{\RRR}{\varrho}

\title{Reduction of Magnetic Noise in Atom Chips by Material Optimization}

\author{V. Dikovsky$^1$, Y. Japha$^1$, C. Henkel$^2$, R. Folman$^1$ \\
{$^1$\it Ben Gurion University of the Negev, Be'er Sheva 84105, Israel} \\
{$^2$\it Institut f\"ur Physik,  Universit\"at Potsdam, D-14469
Potsdam, Germany}}

\maketitle

\begin{abstract}
We discuss the contribution of the material type in metal wires to
the electromagnetic fluctuations in magnetic microtraps close to
the surface of an atom chip. We show that significant reduction of
the magnetic noise can be achieved by replacing the pure noble
metal wires with their dilute alloys. The alloy composition
provides an additional degree of freedom which enables a
controlled reduction of both magnetic noise and resistivity if the
atom chip is cooled. In addition, we provide a careful re-analysis
of the magnetically induced trap loss observed by Yu-Ju Lin {\it
et al.} [Phys. Rev. Lett. {\bf 92}, 050404 (2004)] and find good
agreement with an improved theory.
\\[5ex]
PACS numbers:
\\
39.25.+k Atom manipulation
\\
72.15.-v Electronic conduction in metals and alloys
\\
07.50.-e Electrical and electronic instruments and components
\\
03.75.-b Matter waves

\end{abstract}

\section{Introduction}

Much progress has been made during the last 2-3 years in the
technology of atom chip fabrication
and the experimental control and imaging of ultracold neutral
atoms trapped near metallic and dielectric surfaces. This resulted
in a significant improvement in the study of atom-surface
interactions and provided reliable and reproducible information
about the behavior of the atomic clouds trapped at different
heights above the chip surface \cite{Fortagh}-\cite{Treutlein}. In
magnetic traps close to the surface of the chip, atoms experience
magnetic fluctuations which enhance harmful effects such as
spin-flip induced trap loss, heating and decoherence. Good
agreement was found between experimental measurements of these
effects and their theoretical predictions presented in references
\cite{Varpula}-\cite{Henkel4}. A complete theoretical treatment of
magnetic fluctuations close to metallic surface is based on
summing the contribution of thermally occupied modes of the
electromagnetic field in the presence of the metallic surface.
This approach provides a manageable calculation of the magnetic
fluctuations close to an infinite planar
surface~\cite{Varpula}-\cite{Henkel1}. A different approach (the
quasi-static approximation) assumes that at low transition
frequencies, magnetic fluctuations are instantaneously generated
by thermal electric current fluctuations in the metal
\cite{Henkel2}. This approach is not restricted to planar surfaces
and is therefore suitable for calculating magnetic noise on an
atom chip with complex structures. Both theoretical approaches
predict the same dependence of the magnetic noise on the
temperature and resistivity of the metal chip elements 
in the limit of a large penetration depth (low transition frequencies),
although some differences
occur in the dependence on the chip geometry. For a detailed
review of the relevant equations, see the paper by C. Henkel in
this Special Issue \cite{HenkelS}.

Atom chip surfaces have to meet stringent requirements and are
conventionally made of copper and gold. Both metals are convenient
in view of their ability to hold high current densities and in
view of available technology (ease of evaporation, electroplating,
and etching), gold being somewhat preferable due to its high
resistance to corrosion. Furthermore, it appeared to be common
wisdom that as the surface induced noise was a function of the
ratio $T/\rho$ (quasi-static approximation) between the surface
temperature and its resistivity, there is no point in cooling the
surface, since $\rho$ usually scales linearly with temperature.

In this work, we discuss the possibility of minimizing the surface
induced noise by a proper choice of the materials and their
temperature. As an example, we discuss several materials (normal
metals or alloys) and convenient temperature regimes in which one
may achieve both minimal resistivity and minimal magnetic noise.
Experimental data suggesting that this is indeed possible is
provided by the large difference in the scattering rates observed
for atom clouds trapped near different metal surfaces
\cite{Harber}.

\section{Material dependence of magnetic noise}

Our analysis is based on the theory of loss and heating for
magnetically trapped atoms near metallic surfaces developed in
\cite{Henkel1} and \cite{Henkel2}. Magnetic trapping of neutral
atoms is limited by transitions into untrapped Zeeman levels. The
transition rate $\Gamma_{0\rightarrow f}$ from the trapped state
$|0\rangle$ into an untrapped state $|f\rangle$ is proportional to
the magnetic field fluctuations in the
 following way \cite{Henkel1}:
\begin{equation}
\Gamma_{0\rightarrow f}=\sum_{i,j}\frac{\langle 0| \mu_i|f\rangle\langle f| \mu_j|0\rangle}{\hbar^2}
S_B^{ij}(\omega_{f0})
\label{Gamma_0f}
\end{equation}
where $\mu_i$ and $\mu_j$  ($i,j=1,2,3$) are the projections of
the atomic magnetic moment on the main axes and
$S^{ij}_B(\omega_{f0})$ is the spectral density of the magnetic
noise at the transition frequency $\omega_{f0}$  (for spin-flip
transitions, $\omega_{f0}$ is the Larmor frequency of the atomic
spin in the center of the trap). The spectral density $S_B^{ij}$
is related to the two-point correlation function of the magnetic
field fluctuations:
\begin{equation}
\langle B^*_i({\bf x},\omega)B_j({\bf
x},\omega')\rangle=2\pi\delta(\omega-\omega')S_B^{ij}({\bf
x},\omega)
\end{equation}
where ${\bf x}$ is the location of the trap center. Typically, one
is in the stationary limit of very low transition frequencies,
where the wavelength of the radiated field is much larger than the
dimensions of the system and
the penetration depth of the electromagnetic field into the
metal is much larger than the thickness of the metal structures.
The spectral density of the magnetic
noise is described in this regime by the simple product of a factor,
which is material-dependent, and a geometrical tensor $Y_{ij}$
\cite{Henkel2}:
\begin{equation}
S_B^{ij}({\bf x}, \omega) = \frac{\mu_{0}^2 k_B T}{4\pi^2
\rho}Y_{ij}({\bf x}) \label{SBij}
\end{equation}
where $k_B$ is the Boltzman constant, $\mu_0$ is the vacuum
permeability, $T$ is the temperature of the metal, and $\rho$ its
static resistivity. The geometrical tensor $Y_{ij}({\bf x})$ at
the trap center is given by (note that the factor
$1/2$ is erroneously missing in Eq.(A.6) of Ref.~\cite{Henkel2}):
\par
\begin{eqnarray}
Y_{ij}&=& {\rm tr}\{X_{ij}\}\delta_{ij}-X_{ij} \nonumber \\
X_{ij}&=& \frac{1}{2}\int d^3 {\bf x}'\frac{({\bf x}-{\bf x}')_i
({\bf x}-{\bf x}')_j}{({\bf x}-{\bf x}')^6} \label{Yij}
\end{eqnarray}
where the integration is done over the volume of the metal element.

The spectral density of the magnetic noise in Eq.~(\ref{SBij}) is
actually independent of the transition frequency and gives rise to
a scattering rate which is proportional to the ratio $T/\rho$.
This result of the quasi-static approximation is valid whenever
the skin depth $\delta=\sqrt{2\rho/\mu_0\omega_{f0}}$ is much
larger than both the distance $d$ of the trap from the metal
structure and the thickness $t$ of the metal structure elements.

In the opposite limit, when $\delta$ is very small, e.g. as the
resistance drops upon cooling, the penetration depth is reduced
according to $\delta \propto \rho^{1/2}$, and the scaling laws
with $\rho$ and distance change. For example, above a metallic
half-space, the quasi-static approximation ($d \ll \delta$)
results in a scaling of $T / (\rho d )$ \cite{Varpula,Nenonen,Henkel2} while for $d \gg \delta$
one finds a scaling of $T \rho^{1/2} / d^4$ \cite{Sidles,Henkel1}
(see also a discussion in \cite{HenkelS,Scheel}).

An expression for the magnetic noise in the intermediate range
where the skin depth is comparable to the thickness or the
distance, exists only
for planar (possibly layered) structures in the form of interpolation
formulas \cite{Varpula,Sidles} or integrals to be computed numerically
(see a review in
Ref.~\cite{HenkelS} and recent work in Refs.~\cite{Scheel,Zhang}).
For an infinitely thick metallic layer,
Ref.~\cite{Henkel1} suggests an interpolation
factor of $\left(1+2d^3/3\delta^3\right)^{-1}$ which should
multiply the expression in Eq.~(\ref{SBij}). This factor was shown
experimentally to give the right trends, but is not very accurate
at intermediate distances $d\sim \delta$, where the integral
expression gives a much better agreement with
experiment~\cite{Harber}. Its important to note that in the regime
of a short skin depth, $d>\delta$, the magnetic noise power is
actually reduced compared to the scaling law $T/\rho$. More
specifically, the large-distance scaling of $T \rho^{1/2}$ now
implies a strong noise reduction upon cooling. In this work, we
will mainly deal with the regime where the factor $T/\rho$ holds.
However, when describing our conclusions for this regime, we will
also define its validity borders and address the consequences of
entering the short skin depth regime.

Taking gold at room temperature as our standard (and using similar
notation as in \cite{Henkel2}), we may write a simplified
expression for $\Gamma$:
\begin{equation}
\Gamma\approx 57 s^{-1}\frac{(T/300K)}{(\rho/\rho_{{\rm Au},300K})}\sum_{ij}\frac{\mu_i\mu_j}{\mu_B^2}
(Y_{ij} \times 1\mu m)
\label{Gamma_st}
\end{equation}
where $\mu_B$ is the Bohr magneton, $\rho_{{\rm Au},300K}$  is the
resistivity of gold at room temperature and $\mu_i,\mu_j$ are
matrix elements of the atomic magnetic moment between relevant
states. To avoid complications of magnetic permeability and
hysteresis of the chip, we consider here only nonmagnetic metals
(having no long-range magnetic order). The resistivity of these
metals is essentially a sum of two contributions $\rho = \rho_{0}
+ \rho_{\rm ph}$: a temperature independent residual resistivity
$\rho_0$, due to scattering of charge carriers by crystal defects
and impurities and a phonon contribution $\rho_{\rm ph}$, which can be
described in the well known Bloch-Gruneisen approximation as
\cite{Ziman}:
\begin{equation}
\rho_{\rm ph}=A
\left(\frac{T}{\Theta}\right)^5\int_0^{\Theta/T}\frac{z^5e^z}{(e^z-1)^2}dz
\label{rho_p}
\end{equation}
where $A$ is constant and $\Theta=\hbar\omega_{\rm ph}/k_B$ is the Debye
temperature, $\omega_{\rm ph}$ being the phonon cutoff frequency. The
asymptotic temperature dependence of the phonon resistivity at
very low temperatures ($T\ll\Theta$) is given by $\rho_{\rm ph} \sim
T^5$, while it becomes linear $\rho_{\rm ph} \sim T$ at high temperatures
($T\gg \Theta$).
One can then identify three regimes for the
behavior of the ratio $T/\rho$ that provides upper limits for the
magnetic noise, as mentioned above:
\begin{itemize}
\item[(i)] at low temperatures the metal resistivity is dominated by the
temperature independent residual resistivity $\rho_0$ and $T/\rho$
will decrease linearly when reducing the
temperature.
\item[
(ii)]    in an intermediate regime where the phonon resistivity
dominates, such that $\rho\gg\rho_0$, but the temperature is less
than the Debye temperature $\Theta$ or comparable to it, the resistivity
grows nonlinearly with temperature ($\rho_{\rm ph}
\sim{T^{\alpha}}$, with $\alpha>1$) and $T/\rho$
increases with reducing temperature.
\item[
(iii)]   at high temperatures the resistivity becomes linear in
$T$ and $T/\rho$  will be temperature independent.
\end{itemize}

For real conductors the residual resistance ratio $\RRR\equiv
\rho(300 {\rm K})/\rho_0$ (used as a quality parameter) has a wide
variation due to its dependence on the purity level, as well as
the thermal and mechanical treatment of the metal. Conducting
elements on an atom chip are usually fabricated as a film and
thereby may have significant residual resistivity relative to bulk
metals. Lattice and surface disorder in metal films, which is
mainly due to grain structure and gas absorption, results in the
increase of $\rho_0$ \cite{Wissmann}.

For the comparison of the magnetic noise produced by different
materials we need not specify the geometry of the conductor, which
is all contained in the tensor $Y_{ij}$. In accordance with
Eq.~(\ref{Gamma_st}), we can formulate the optimization criteria
as the following: the material used for the chip wires should have
both the resistivity and the ratio $T/\rho$ smaller than the gold
standard at $T=$300 K, in a convenient temperature interval.

\section{Pure metals}

\begin{figure}[h]
\includegraphics[width=0.8\textwidth] {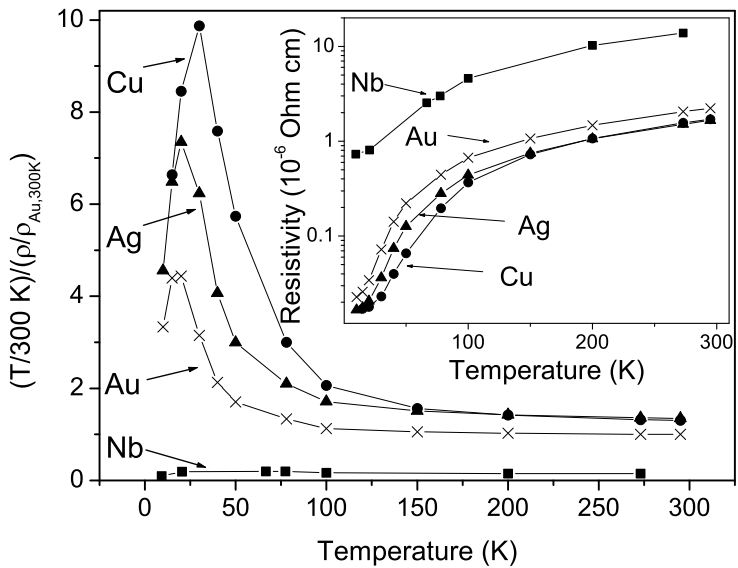}
\caption{Temperature dependence of the normalized magnetic noise
calculated using Eq.~(\ref{Gamma_st}) for wires made of copper
(circles), silver (triangles) gold (crosses) and niobium
(squares). The noise is normalized to the value for gold at
$T=$300 K. Inset: temperature dependence of the resistivity
(extracted from \cite{Malkov}). } \label{Fig1}
\end{figure}

Pure metals usually have high values of $\RRR$, such that for
$10\,{\rm K} < T < 300\,{\rm K}$, their behavior follows that of
the intermediate regime defined above, where the magnetic noise
increases with cooling. This situation is illustrated in
Fig.~\ref{Fig1}, where the normalized magnetic noise of copper,
silver, gold and niobium is presented as a function of
temperature. The values are normalized to the magnetic noise level
for gold at $T=$300 K. Respective resistivity behavior is shown in
the inset. The data for the ideal (phonon) resistivity of the
metals were extracted from \cite{Malkov}. We assume $\rho_0$ to be
1\% of the room temperature resistivity (such values are realistic
for high quality films of 2-3$\mu$m thickness ~\cite{Bassewitz,RRR}). One
should note that the latter number is highly sensitive to the
details of the fabrication process, including material purity,
film thickness, granular structure and surface specularity. As can
be seen from Fig.~\ref{Fig1}, the magnetic noise from the studied
noble metals is not reduced by cooling but rather exhibits a
pronounced peak in the temperature range 20-50 K.

As mentioned above in the discussion following Eq.~(\ref{SBij})
and (\ref{Yij}), we need to define the validity borders of the
$T/\rho$ approximation. As an example, we consider a trap with a
spin-flip transition frequency $\omega_{f0} \approx 2\pi\times
0.79$MHz ($B_0$=0.57G) near a copper-made conductor, as in the
experiment described in Ref.~\cite{Lin}. The skin depth for copper
with $\RRR$=100 is $\delta=74\mu$m at room temperature
($\rho=1.7\mu\Omega\cdot$cm), $\delta=25.3\mu$m at liquid nitrogen
temperature ($\rho=0.2\mu\Omega\cdot$cm) and $\delta=7.4\mu$m at
liquid helium temperature ($\rho=0.017\mu\Omega\cdot$cm).
Following the explanation given above, our analysis of pure metals
at low temperatures  will be restricted to trap distances from the
surface of less than $\delta$, namely, less than 7$\mu$m at liquid
helium temperatures. Some corrections will be needed for distances
close to 7$\mu$m and a separate treatment will be needed for
longer distances. For higher temperatures, or for alloys with
higher resistivity as in the next section, or for lower transition
frequencies such as transitions between motional levels in
magnetic traps (heating), the following analysis will be valid for
distances of at least several tens of microns.

\begin{figure}
\includegraphics [width=0.8\textwidth] {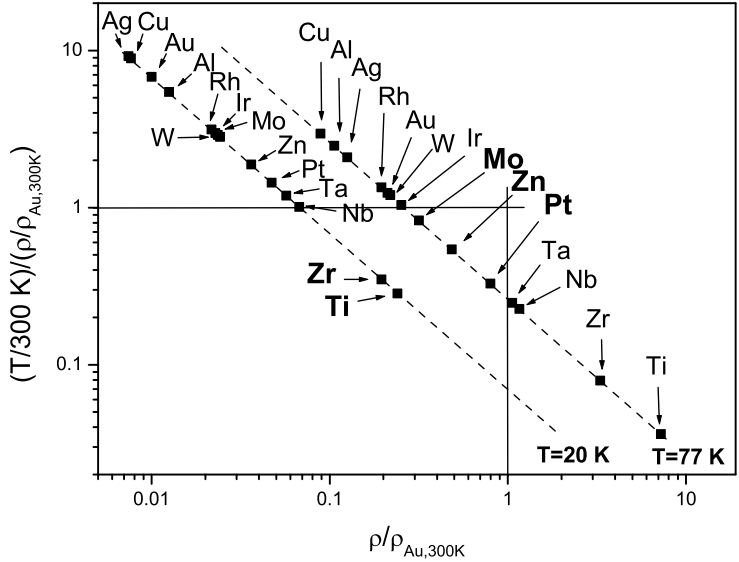}
\caption{Dependence of the $T/\rho$ ratio (normalized to its value
for gold at 300\,K) on the normalized resistivity at $T=$77 and
20 K for various metals. The dashed lines correspond to the
theoretical dependencies of the noise on the resistivity for the
given temperatures (see Eq.~(\ref{SBij})). Metals having qualities better
than those of gold at 300 K, are indicated in bold letters.}
\label{Fig2}
\end{figure}

 In order to broaden the search we have performed the magnetic noise
calculation for 14 metals having no magnetic ordering: Al, W, Au,
Ir, Cu, Mo, Nb, Pt, Rh, Ag, Ta, Ti, Zn, Zr. The resistivity data
for the metals were extracted from \cite{Malkov}. It is convenient
to plot the normalized noise vs. the normalized resistivity at
constant temperature. In accordance with Eq.~(\ref{Gamma_st}),
this plot gives a straight line with unit negative slope if
plotted in a double logarithmic scale. The results of the analysis
for the temperature of $T=$77 and 20 K are displayed in
Fig.~\ref{Fig2}. Both the noise and resistivity data are
normalized to those of gold at $T=$300 K (thus, points appearing
in the area restricted by the solid lines correspond to values
better than those of gold at room temperature). We may observe
that at least three metals (molybdenum, zinc and platinum) at
liquid nitrogen temperature and two metals (zirconium and
titanium) at $T=$20 K have better characteristics than those of gold
at $T=$300 K. For example, Ti-made wires cooled down to 20 K
will have a factor three reduction in the noise and four-fold
reduction in the resistivity relative to gold at room temperature.

Finally, let us note that in this analysis we are considering only
the magnetic noise reduction due to the $T/\rho$ ratio appearing
in Eq.~(\ref{SBij}).
The geometrical factor $Y_{ij}$ may also contribute to noise
reduction because it depends on wire thickness. As the resistivity
drops with cooling, wire thickness may be reduced while
maintaining the same current. However, this may give rise to
problems such as enhanced relative 
surface roughness, which may in turn enhance harmful effects such
as fragmentation. 
In addition, thickness reduction is limited as
thinner films have an increased resistivity.

\section{Alloys}

\begin{figure}
\includegraphics [width=0.8\textwidth] {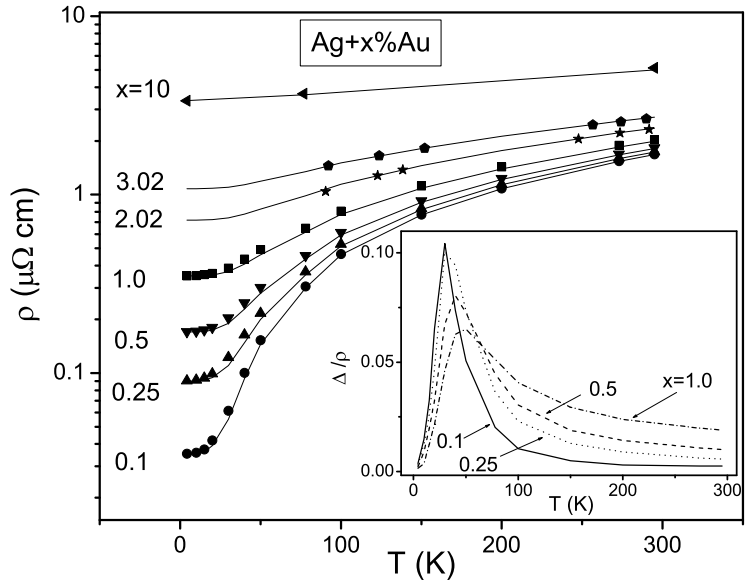}
\caption{Measured resistivity of Ag-Au alloys with different
concentrations as a function of temperature \cite{Dugdale,Linde1,
Davis}. Solid lines are the predicted values according to the
Matthiessen's rule. Inset: temperature dependence of the ratio of
the deviation $\Delta(x,T)$ from the Matthiessen's rule to the
total Ag-Au alloy resistivity for small Au concentrations.}
\label{fig:rhoTx}
\end{figure}

More promising materials for a cooled atom chip could be found
among nonmagnetic dilute metal alloys, whose residual resistivity
is usually higher than in pure metals. In alloys, such residual
resistivity may be set to a desired value by adjusting the
material composition. The resistivity of a dilute binary alloy is
given by
\begin{equation}
\rho(x,T)=\rho_0(x)+\rho_{\rm ph}(T)+\Delta(x,T),  \label{rhoTx}
\end{equation}
where the residual resistivity $\rho_0$ is proportional to the
concentration $x$ of the solute metal and $\rho_{\rm ph}$ is the
phonon resistivity of the solvent metal, which is not changed by
the addition of the solute metal, according to the well known
Matthiessen's rule. The last term in Eq.~(\ref{rhoTx}),
$\Delta(x,T)$, is a correction to the Mathiessen's rule, which was
intensively investigated in the literature (see, for example,
\cite{Bass}). This deviation represents a change in the
temperature dependence of the resistivity at low temperatures and
moderate solute concentrations and its measurement is important
for the study of scattering mechanisms in metals and alloys (for
more details see Refs.~\cite{Bass}-\cite{Campbell}).
However, its value relative to the total resistivity
is small. Figure~\ref{fig:rhoTx} shows the measured values of
resistivity for Ag-Au alloys at different concentrations and
temperatures. The inset shows that the deviation $\Delta$ from the
Matthiessen's rule is less than 10\% for Au concentrations
between 0.1\% and 10\%. At concentrations as high as 2\% or more
it is negligible.
In the following analysis we will therefore
neglect the correction term $\Delta$ and assume that the
Matthiessen's rule is a good approximation for the calculation of
the magnetic noise.

\begin{figure}
\includegraphics [width=0.8\textwidth] {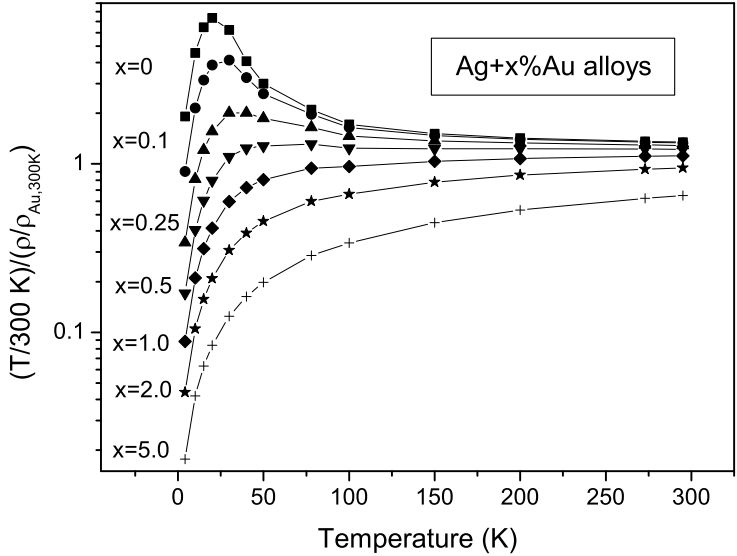}
\caption{Temperature dependence of the ratio $T/\rho$ (normalized
to its value for gold at 300 K) for silver and its alloys with
gold: pure silver (squares), and with 0.1\% gold (circles), 0.25\%
(triangles), 0.5\% (inverted triangles), 1\% (diamonds), 2\%
(stars), and 5\% (crosses). The $\rho(T)$ dependence for alloys
was calculated using the residual resistivity data given by
\cite{Dugdale,Linde}. Note the difference compared to Fig. 1.}
\label{Fig3}
\end{figure}

If the solute concentration is not very high (up to $x=$10-15\%,
where ordering processes may start changing the behavior) the
residual resistivity of the alloy $\rho_0(x)$ increases linearly
with increasing component ratio \cite{Bass,Dugdale,Fairbank}. This
concentration dependent residual resistivity provides an
additional degree of freedom which can be controlled in order to
obtain the desired magnetic noise and resistivity of the chip
material. The concentration of the solute should not be too large.
In accordance with our optimization criteria, the alloy is
effective as a chip wire material if its total resistivity does
not exceed gold resistivity at room temperature. In
Fig.~\ref{Fig3} the temperature behavior of the normalized ratio
$T/\rho$ is shown for a set of dilute Ag-Au alloys as well as for
pure silver. The temperature dependencies $\rho(T$) for the alloys
were calculated using Eq.~(\ref{rhoTx}). The phonon resistivity
data of silver ($\rho_{{\rm ph,\,Ag}}(T)$) were extracted from
\cite{Malkov}, and the experimental values of the residual
resistivity ($\rho_0(x)$) for different gold concentrations $x$
were extracted from references \cite{Dugdale} and \cite{Linde}.

\begin{figure}
\includegraphics [width=0.8\textwidth] {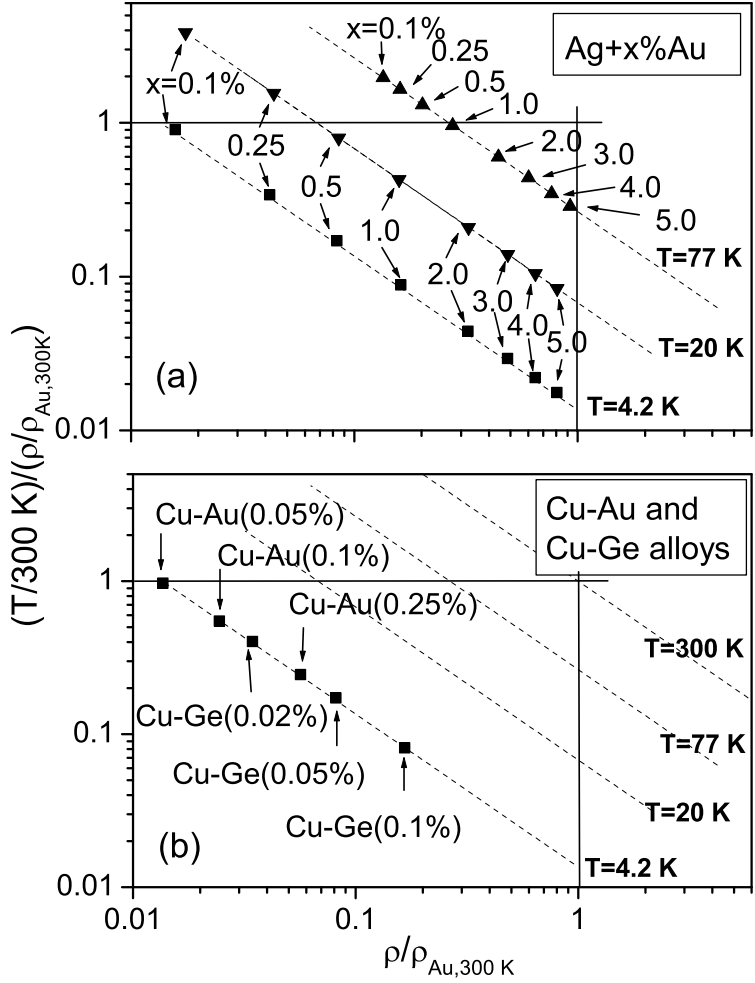}
\caption{Normalized noise ($T/\rho$) and resistivity
characteristics for dilute alloys. The values for $T/\rho$ and
resistivity are normalized to those of gold at 300 K. (a) Ag-Au
alloys, (b) Cu-Au and Cu-Ge alloys. For (b) the solute
concentration is given in brackets. The residual resistivity data
for alloys are extracted from ~\cite{Dugdale} and ~\cite{Linde}.}
\label{Fig4}
\end{figure}

From Fig.~\ref{Fig3} one may conclude that the low temperature
maximum in $T/\rho$, which is intrinsic for pure metals
(Fig.~\ref{Fig1}), disappears with increasing $x$. This behavior
is general for most of the dilute alloys. In particular, it is
valid for the alloys we analyze in the following. In
Fig.~\ref{Fig4} we plot the normalized noise vs. normalized
resistivity characteristics of the alloys Ag-Au, Cu-Au and Cu-Ge.
As an example, the cooling of wires made of the alloy Ag+5\%Au
down to $T=$77 K can reduce the noise by a factor of 5 relative to
the level of pure silver at room temperature, and by a factor 3
relative to gold. The latter factor is almost the same as that
achieved, for example, for Pt in Fig.~\ref{Fig2}, but due to the
unique alloy phenomena described above (Fig.~\ref{Fig3}), a
pronounced difference will appear between Pt and Ag+5\%Au as the
temperature is lowered further.

According to Eq.~(\ref{SBij}), the product of the magnetic noise
and the resistivity is linearly proportional to the temperature
and independent of the material (as indicated by the sloped
straight lines in Fig.~\ref{Fig4}). A cooling procedure would
reduce the magnetic noise relative to the magnetic noise of gold
at room temperature by a factor $(T/300 K)\cdot(\rho_{{\rm Au},300
K}/\rho_{Alloy}(T))$. It is possible to choose the alloy
concentration such that $\rho_{\rm Alloy}(T)=\rho_{{\rm Au},300
K}$, so that the resistivity will be kept at the gold standard but
the magnetic noise will be reduced by a factor of $T/$300K. Hence,
the noise of an alloy-made conductor could be reduced relative to
the gold standard maximally by 4 times when cooling to liquid
nitrogen temperature ($T=$77 K) and by 70 times by cooling to
liquid helium temperature ($T=$4.2 K). This ratio of noise
reduction is reached on the boundary of the optimization area. In
particular, for the alloys Ag-Au, Cu-Au and Cu-Ge, the solute
concentrations on this boundary equal 6\%, 4.5\% and 0.52\%
respectively (for $T=$4.2 K). The values were obtained from the
$\rho_0(x)$ dependencies presented in \cite{Dugdale,Davis}.

Thus, the main conclusion of this paper is that a significant
simultaneous reduction of magnetic noise and resistivity is
possible with both pure metals and alloys at 77K, and may be
reduced even further by making use of alloys and further cooling.
We note that a further advantage of wires made of alloys relative
to pure metals, may perhaps be found in the fact that their
resistivity is less sensitive to temperature fluctuations, since
it is mainly due to the residual resistivity. This fact may
contribute to current stability under temperature variations in
space (along the wire) or time.

\begin{figure}
\includegraphics [width=0.8\textwidth] {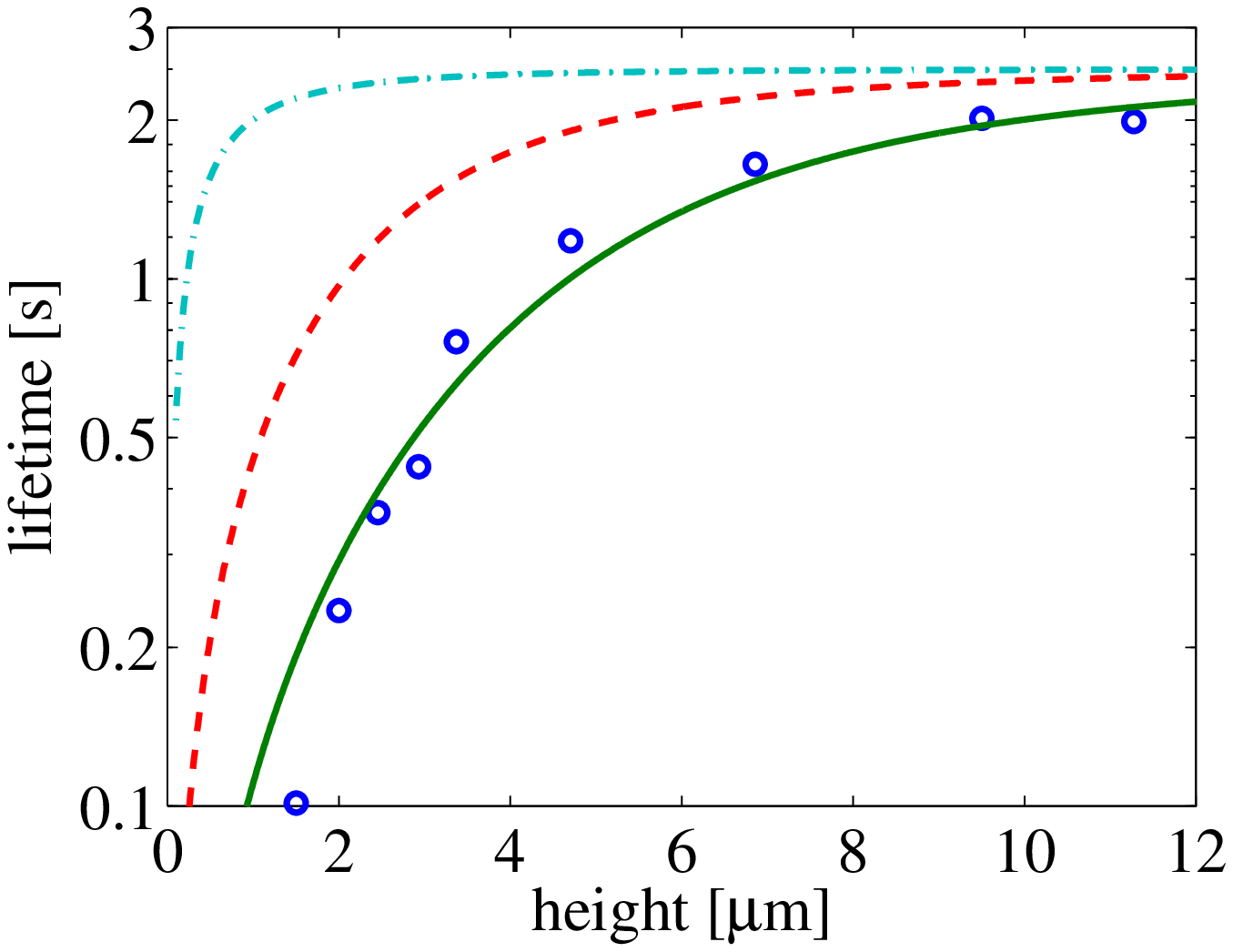}
\caption{Comparison of trapping lifetimes of $^{87}$Rb atoms above
a copper wire on an atom chip~\cite{Lin} with a theoretical
calculation (solid line). Predicted lifetimes are also shown for a
similar wire made of an alloy of Ag with 6\% Au content, cooled
down to $T=$77K (dashed) and 4.2K (dash-dotted). In both cases the
resistivity was taken equal to gold resistivity at $T$=300 K,
$\rho_{\rm Au}=2.21\mu\Omega \cdot$cm (same penetration depth).
According to Fig.
~\ref{Fig4}a, the needed gold concentrations are estimated by
$x=5.4$\% and $x=6$\%, respectively. Note that our calculation
differs from the one made in~\cite{Lin} (see Appendix A). Van der Waals
forces are not taken into account. Let us also mention that the
maximum noise reduction factor of 70 noted in the text, is not
visible, due to the effect of the technical noise and/or
background gas collisions limiting the lifetime in this experiment
to a maximum of $\tau_{\rm tech}=2.5\,s$~\cite{LinPrivate}.}
\label{Fig5}
\end{figure}

In order to illustrate the effect of noise reduction on the
lifetime of trapped atoms we present the predicted lifetime of
$^{87}$Rb close to a metal wire similar to that used in
Ref.~\cite{Lin}. Fig.~\ref{Fig5} shows the experimental results
together with a theoretical curve simulating the experiment with
the copper wire at $T= $400 K ($\rho_{\rm
Cu}=2.64\mu\Omega\cdot$cm). Details of the calculation are given
in Appendix A (note that our calculation is different from that
made in~\cite{Lin}). No fitting parameters were used, except for
assuming a distance independent scattering rate of 0.4
s$^{-1}$\cite{Lin,LinPrivate}. The predicted lifetimes for a
similar wire made of the alloy Ag+6\%Au, and cooled down to $T=
$77K or 4.2K, are also presented. They show that lifetimes can be
increased by more than an order of magnitude. When cooling down to
4.2K, the atomic lifetime is close to the limit set by (distance
independent) background scattering even when the atoms are held 1$\mu
m$ from the wire.

\begin{figure}
\includegraphics[width=0.8\textwidth] {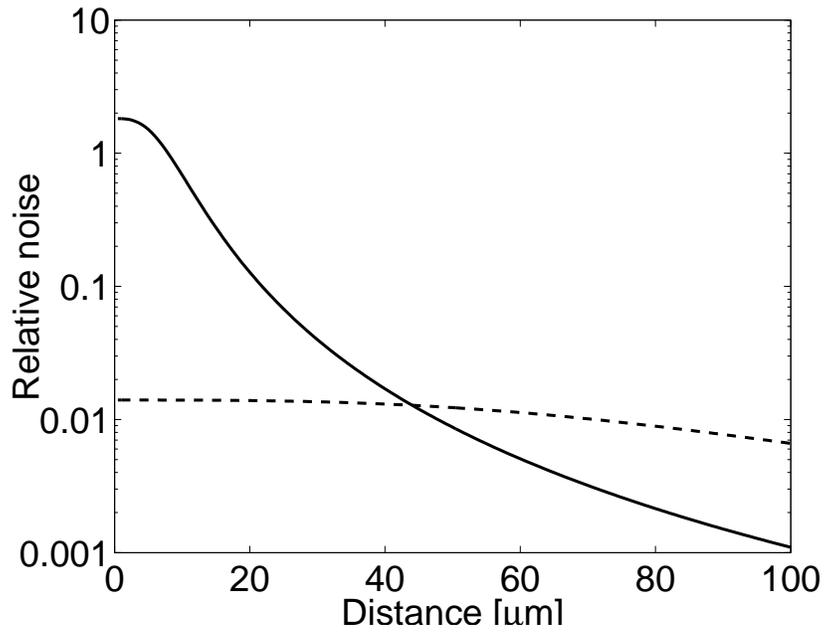}
\caption{Magnetic noise relative to the room temperature gold
standard for an Ag-Au alloy (dashed line) at a temperature of 4.2
K and for a pure copper wire at 4.2 K (solid line) as a function
of distance from the chip. The noise was calculated based on the
interpolation formula of Ref.~\cite{Henkel1} (see text). The
penetration depths used for copper and the Ag-Au alloy are
$\delta=7.4\mu$m and $\delta=84\mu$m, respectively.
\label{reduction}}
\end{figure}

As suggested by Fig.~\ref{Fig1}, the magnetic noise will not be
reduced by cooling the copper wires, as long as it obeys the
$T/\rho$ law. However, as the temperature becomes smaller, one
approaches the regime where the penetration depth $\delta$ gets
shorter than characteristic scales like the trap height $d$, and
the $T/\rho$ law must be amended. As mentioned before, in the
simple case of an infinitely thick metal layer, the noise power
can be multiplied by the interpolation factor
$(1+2d^3/3\delta^3)^{-1}$ to get the scaling for short skin depth
\cite{Henkel1}. At large distances, the $T/(\rho d)$ law is thus
replaced by the asymptotic law
$\frac{3}{2}(2/\mu_0\omega_{f0})^{3/2}T\sqrt{\rho}/d^4$ with a
more favorable scaling with temperature. In Figure~\ref{reduction}
we compare the noise (relative to the standard of gold at room
temperature) for a Ag-Au alloy and for a copper wire at 4.2 K, as
a function of distance. It is seen that the noise from a cooled
copper wire drops below the noise from a wire made of an alloy
only at $d>45\mu$m. This value is expected to be even bigger if a
wire thickness $t<\delta$ would be taken into account. We recall
that the interpolation factor mentioned above actually
overestimates the magnetic noise power, as illustrated in
Ref.\cite{Henkel1} and in Ref. \cite{Harber} by comparing to
numerical calculation and experimental data, respectively. The
maximal observed discrepancy is about a factor of 3 at $d=\delta$.
On the other hand, one can expect an increase of the pure metal
$\delta$ at cold temperatures, due to the fact that typical pure
metal wires have larger residual resistivities than the numbers
used in our calculation. For example, Ref.~\cite{Lin} calculated
$\delta =103\,\mu{\rm m}$ while our calculation uses $\RRR=100$
which leads to $\delta=92\,\mu{\rm m}$. In summary, it follows
that as far as magnetic noise is concerned, cooled wires made of
alloys are preferable over cooled wires made of pure metals over a
large range of distances from the atom chip.

Finally, we consider in Appendix B the feasibility of cooling
current carrying wires to low temperatures. We analyze the
expected heating due to the ohmic resistance and find that for
considerable current densities, the temperature rise in the wires
does not significantly alter the results presented in this paper.

\section{Conclusions}

We have analyzed the dependence of magnetic fluctuations near the
surface of conducting elements on an atom chip, on the material
type and the temperature. We have shown that at low temperatures a
significant reduction of the noise can be achieved by making the
conducting wires from dilute alloys of noble metals.

This improvement is due to the large value of the temperature
independent residual resistivity of the alloys compared to pure
noble metals, which is controllable by smoothly changing the
solute component concentration. At least three important
advantages of cooling wires made of alloys could be underlined:
(i) reduction of the magnetic noise, thereby decreasing loss rate,
heating and decoherence; (ii) possible reduction of the
resistivity of the wires, and (iii) enhancement of temperature
stability.
With the
data presented in this work, it is possible to identify alloys
that reach a minimum noise level that is set by that of
the expected technical noise.

Finally, we note that for optimal noise reduction, a complete
optimization is required for every specific situation. Such an
optimization should take into account not only the $T/\rho$ ratio
analyzed in this work, but also the geometrical factor, which
includes important parameters such as wire thickness, and trap
height (see the discussion in Ref. \cite{Zhang} for flat wires).
We note that the geometrical parameters also define the validity
borders of the $T/\rho$ approximation, and corrections have to be
applied when the penetration depth becomes comparable to wire
thickness or trap height.

\section*{Acknowledgments}
We thank Yu-Ju Lin and Joerg Schmiedmayer for useful discussions.
This work was supported by the European Union FP6 "atomchip"
collaboration (RTN 505032), the German Federal Ministry of
Education and Research (BMBF) through the DIP project, and the
Israeli Science Foundation. C.H.\ acknowledges financial support
from the European Commission (contracts FASTNet HPRN-CT-2002-00304
and ACQP IST-CT-2001-38863) and thanks Jens Eisert and Martin
Wilkens for contributing to a stimulating environment.

\appendix

\section{Comparison between predicted and measured magnetic noise for
thin rectangular wires}

For calculating the magnetic noise close to a rectangular  metal
wire of width $w$ and thickness $t$ we have calculated the tensor
$Y_{ij}$ of Eq.~(\ref{Yij}) for a slab extending from $y'=-\infty$
to $y'=\infty$, from $x'=-w/2$ to $x'=w/2$ and from $z'=-t$ to
$z'=0$. The results of the integration for nonzero elements
$Y_{ij}$ are:
\begin{eqnarray}
Y_{11} & = & \frac{3}{8}{\rm tr}\{Y_{ij}\}+\frac{\pi}{16}\left[\left[\frac{v}{u}\frac{1}{\sqrt{u^2+v^2}}
\right]_{u=-w/2-x}^{w/2-x}\right]_{v=-z-t}^{-z} \\ \label{Y11}
Y_{22} & = & \frac{3}{8}{\rm tr}\{Y_{ij}\}=-\frac{3\pi}{16}\left[\left[\frac{\sqrt{u^2+v^2}}{uv}
\right]_{u=-w/2-x}^{w/2-x}\right]_{v=-z-t}^{-z} \\
Y_{33} & = & \frac{3}{8}{\rm tr}\{Y_{ij}\}+\frac{\pi}{16}\left[\left[\frac{u}{v}\frac{1}{\sqrt{u^2+v^2}}
\right]_{u=-w/2-x}^{w/2-x}\right]_{v=-z-t}^{-z} \\
Y_{31} & = & \frac{\pi}{16}\left[\left[\frac{1}{\sqrt{u^2+v^2}}
\right]_{u=-w/2-x}^{w/2-x}\right]_{v=-z-t}^{-z} \label{Y31}
 \end{eqnarray}
where we have used the definition
\begin{equation}
\left[\left[f(u,v)\right]_{u=a}^{b}\right]_{v=c}^{d}\equiv f(a,c)-f(a,d)-f(b,c)+f(b,d).
\end{equation}
In Ref.~\cite{Lin} the lifetime of trapped atoms was measured
above the center of a rectangular copper wire of thickness
$t=2.15\mu m$ and width $w=10\mu m$. The temperature of the wire
was estimated as 400 K ($\rho_{\rm Cu}=2.64\mu\Omega\cdot$cm). The measurements were performed with
$^{87}$Rb atoms trapped in the hyperfine state
$|F,m\rangle=|2,2\rangle$. The loss process was assumed to be due
to the cascade of transitions $|2,2\rangle \rightarrow |2,1\rangle
\rightarrow |2,0\rangle$. By use of Eq.~(\ref{Gamma_0f}) and
Eq.~(\ref{SBij}) we obtain for this kind of transitions
\begin{equation}
\Gamma_{|F,m\rangle \rightarrow |F,m-1\rangle}=
\frac{\mu_{0}^2\mu_B^2 g_F^2 k_B T}{4\pi^2\rho}\sum_j
Y_{jj}\left|\langle F,m|F_j|F,m-1\rangle \right|^2
\end{equation}
where we have substituted $\mu_j=\mu_B g_F F_j$. The contribution
of off-diagonal terms of $Y_{ij}$ vanishes, given the symmetric
position above the center of the wire.

In Ref.~\cite{Lin}, the total lifetime of the trapped states due
to magnetic noise was approximated to be $\tau_{mag}\approx
\tau_1+\tau_2$, where $\tau_2$ and $\tau_1$, which are the
lifetimes of the respective levels $|2,2\rangle$ and
$|2,1\rangle$, are equal to the inverse of the
transition rates $\gamma_{21} = \Gamma_{|2,2\rangle \rightarrow |2,1\rangle}$
and $\gamma_{10} = \Gamma_{|2,1\rangle \rightarrow |2,0\rangle}$.
A complete approach should take into account
that the transition frequency between the magnetic levels is in
the order of MHz, while the temperature of the surface is several
orders of magnitude larger. This means that the atoms have an
equal probability for transitions from $|2,2\rangle$ into
$|2,1\rangle$ and vice versa. We assume that atoms that decayed
into the untrapped level $|2,0\rangle$ escape immediately from the
trap region. The rate equations are then
\begin{eqnarray}
\frac{d}{dt}P_{|2,2\rangle} &=& -\gamma_{21}(P_{|2,2\rangle}-P_{|2,1\rangle}) \\
\frac{d}{dt}P_{|2,1\rangle} &=& \gamma_{21}(P_{|2,2\rangle}-P_{|2,1\rangle})-\gamma_{10}P_{|2,1\rangle} \\
\frac{d}{dt}P_{|2,0\rangle} &=& \gamma_{10}P_{|2,1\rangle}
\end{eqnarray}
For our calculation, we make use of $|\langle
2,2|F_j|2,1\rangle|^2 = 1$ and $|\langle 2,1|F_j|2,0\rangle|^2 =
3/2$ for $j=x,z$, where we assume the direction of the magnetic
field to be along the wire axis ($y$ direction). We thus have
$\gamma_{10}/\gamma_{21}=3/2$, and the solution to the rate
equations yields
\begin{equation}
P_{mag}=P_{|2,2\rangle}+P_{|2,1\rangle}=\frac{1}{5}\left[6e^{-\gamma_{21}t/2}-e^{-3\gamma_{21}t}\right]
\end{equation}
This gives an effective ($1/{\rm e}$) lifetime of
$\tau=2.364\gamma_{21}^{-1}$.

\begin{figure}
\includegraphics [width=0.8\textwidth] {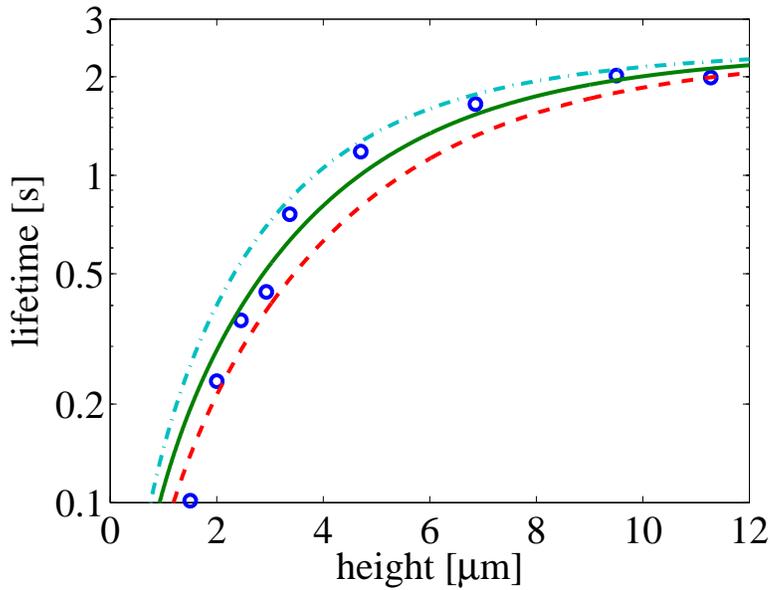}
\caption{The trapping lifetimes of $^{87}$Rb atoms above a wire on
an atom chip for the data presented in Ref.~\cite{Lin}. The plot
presents a comparison between the two calculation methods
presented in the text and the calculation of Ref.~\cite{Lin} (dash-dotted line).
The dashed line
represents the taking into account of transitions in one direction only while the solid
line represents the complete approach.} \label{Fig6}
\end{figure}

In Fig.~\ref{Fig6} we compare this calculation (solid line) to the
simpler model described above where  $\tau_{\rm
mag}=\tau_1+\tau_2$ (dashed line), while using the same
geometrical factors. For completeness, we also present the
calculation of ref.~\cite{Lin} (dash-dotted line), who have used
an interpolated formula for the geometrical factor. The
experimental trap lifetime is restricted by technical noise and/or
background gas collisions to a maximum of $\tau_{\rm
tech}=2.5\,s$~\cite{LinPrivate}. Hence the total trap lifetime is
equal to $\tau_{\rm trap}=(\tau_{\rm mag}^{-1}+\tau_{\rm
tech}^{-1})^{-1}$. The geometrical factors $Y_{ij}$ were
calculated using Eqs.~(\ref{Y11})-(\ref{Y31}). We see that the
complete approach (solid line) gives a better agreement with
experiment than the calculation based on the simplified model
(dashed line). Based on this comparison, we make use of the
complete approach, and present in Fig.~\ref{Fig5} the expected
trap lifetime for the alloy Ag-Au for different cooling
temperatures.

Hopefully, one would be able to construct a more complete theory
that is compatible with the full theory presented in
ref.~\cite{Henkel1}, while still enabling a practical calculation
of finite structures \cite{HenkelS}. We then expect that at low
heights above the surface (smaller than the penetration depth in
the metal), the parallel components $Y_{11}$ and $Y_{22}$ will
decrease by a factor of 1/3, giving rise to an increase of the
calculated lifetime by a factor of 5/3. On the other hand, the
measurements of ref.~\cite{Lin} were performed with a thermal
atomic cloud of temperature 1$\,\mu$K. One then expects that the
cloud expands in the $x,z$ directions to a radius of $\sim
0.5\,\mu$m, so that the loss rate is significantly increased for
the fraction of atoms that are closer to the metal surface. The
deformation of the trapping potential due to atom-surface
interactions of the van der Waals type (not included in
Fig.~\ref{Fig5}) may also contribute to an increased loss rate.
These effects may perhaps explain the noted difference between all
the calculated curves and the experimental results at low heights
above the surface.

\section{Feasibility of cooling current carrying wires}

In Ref.~\cite{Groth} it was shown that the heating of a metal wire
on an atom chip is characterized by two time scales: a fast time
scale related to heat transfer through the contact and/or
insulating layer from the metal into the substrate followed by
slow heating due to heat transfer in the substrate. Heat transfer
through the contact and/or insulating layer depends on the heat
conductivity $k$ of these layers. For a single contact layer of
effective thickness $d_c$ and heat conductivity $\lambda_c$ the
number $k$ is given by $k=\lambda_c/d_c$. For a wire of width $w$,
thickness $h$ and resistivity $\rho(T)$ the equation of heat
transfer is
\begin{equation} hC_W \frac{\partial}{\partial t}\Delta T=h\rho(T)j^2-k\Delta T,
\end{equation}
where $C_W$ is the heat capacity of the metal, $h\rho j^2$ is the
rate of generation of heat per unit wire area and $\Delta T=T-T_0$
is the temperature rise above the substrate temperature $T_0$. To
get a rough estimate of $\Delta T$ after the heating we assume
here that $\rho(T)$ is constant within $\Delta T$ (for instance,
as presented in Fig.~\ref{Fig4}a, the resistivity of our example
alloy Ag+5\%Au hardly changes in the range of 4-77K). We then
obtain a steady state value of $\Delta T$:
\begin{equation} \Delta T=h\rho j^2/k\,. \end{equation}

The process of slow heating was modeled in Ref.~\cite{Groth} in a
two dimensional model with an analytic and numeric solution. It
was found that the slow temperature rise is given by
\begin{equation} \Delta T_s=\frac{hw\rho j^2}{2\pi\lambda}\ln\left(\frac{4\pi^2 \lambda t}
{Cw^2}\right), \end{equation}
where $C$ and $\lambda$ are the heat
capacity and conductivity of the substrate, respectively.

\begin{figure}
\includegraphics [width=0.8\textwidth] {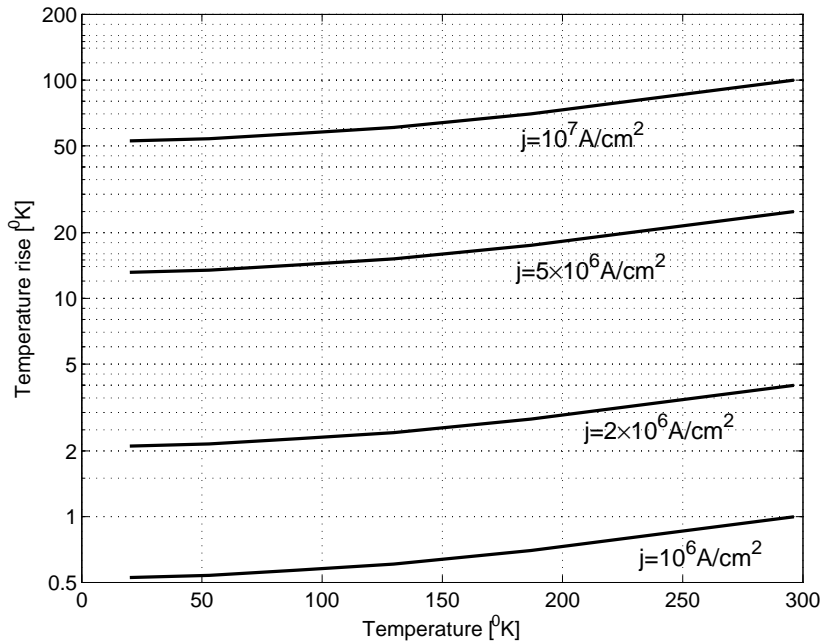}
\caption{Temperature rise in a current carrying wire as a function
of the initial system temperature, and for different current
densities. The wire geometry was taken to be as in
Ref.~\protect\cite{Groth}. The temperature rise includes the fast
rise and a slow rise for 30s.} \label{fig9}
\end{figure}

Typical values of $\Delta T$ as a function of the initial
temperature are given in Fig.~\ref{fig7} for a few values of
current density $j$. The temperature rise contains the fast
heating and slow heating after $30$s. The calculation was done for
the geometry of Ref.~\cite{Groth}, namely $w=5\mu$m, $h=1.4\mu$m
and $\rho=\rho_{\rm Au,300\,K}$. Since the dependence of the heating on
the current density $j$ is quadratic, it follows that if the
temperature of a wire holding a current density of $10^7$ A/cm$^2$
rises by 50 K, then a wire with current density of 10$^6$A/cm$^2$
heats only by 0.5 K.

We may conclude that the reduction of magnetic noise by cooling
will always be limited by wire heating if the current density in
the wire is large. The heating will still be negligible for
current densities of the order of $10^6$A/cm$^2$ or less. One
should note that improved geometries, such as wires buried in the
wafer (thus transferring heat to the wafer through 3 facets), will
further improve the allowed current densities.

\end{document}